\documentclass[aps,prl,superscriptaddress,reprint,showpacs,amssymb,amsmath,floatfix,citeautoscript,longbibliography]{revtex4-2}
 
\usepackage{graphicx,hyperref} 
\usepackage{dcolumn} 
\usepackage{comment}
\usepackage{ulem}
\usepackage[]{color}

\usepackage{tabularx}
\usepackage{multirow}
\newcolumntype{Y}{>{\centering\arraybackslash}X}
\usepackage{ulem}

\usepackage{booktabs}
\usepackage{epsfig}

\begin{document}


\title{Stabilization of U\,5$f^2$ configuration in UTe$_2$ through U\,6d dimers in the presence of Te2 chains}

\author{Denise~S.~Christovam}
\affiliation{Max Planck Institute for Chemical Physics of Solids, N{\"o}thnitzer Stra{\ss}e 40, 01187 Dresden, Germany}

\author{Martin Sundermann}
\affiliation{Max Planck Institute for Chemical Physics of Solids, N{\"o}thnitzer Stra{\ss}e 40, 01187 Dresden, Germany}
\affiliation{PETRA III, Deutsches Elektronen-Synchrotron DESY, Notkestra{\ss}e 85, 22607 Hamburg, Germany}

\author{Andrea~Marino}
\affiliation{Max Planck Institute for Chemical Physics of Solids, N{\"o}thnitzer Stra{\ss}e 40, 01187 Dresden, Germany}

\author{Daisuke~Takegami}
\affiliation{Max Planck Institute for Chemical Physics of Solids, N{\"o}thnitzer Stra{\ss}e 40, 01187 Dresden, Germany}
\altaffiliation{Present address: Department of Applied Physics, Waseda University, Shinjuku, Tokyo 169-8555, Japan}

\author{Johannes~Falke}
\affiliation{Max Planck Institute for Chemical Physics of Solids, N{\"o}thnitzer Stra{\ss}e 40, 01187 Dresden, Germany}

\author{Paulius~Dolmantas}
\affiliation{Max Planck Institute for Chemical Physics of Solids, N{\"o}thnitzer Stra{\ss}e 40, 01187 Dresden, Germany}

\author{Manuel~Harder}
\affiliation{Department of Physics, Universit{\"a}t Hamburg, Notkestraße 9-11, 22607 Hamburg, Germany}
\affiliation{European XFEL GmbH, Holzkoppel 4, 22869 Schenefeld, Germany}

\author{Hlynur~Gretarsson}
\affiliation{PETRA III, Deutsches Elektronen-Synchrotron DESY, Notkestra{\ss}e 85, 22607 Hamburg, Germany}
\affiliation{Max Planck Institute for Solid State Research, Heisenbergstra{\ss}e 1, 70569 Stuttgart, Germany}

\author{Bernhard~Keimer}
\affiliation{Max Planck Institute for Solid State Research, Heisenbergstra{\ss}e 1, 70569 Stuttgart, Germany}

\author{Andrei Gloskovskii}
\affiliation{PETRA III, Deutsches Elektronen-Synchrotron DESY, Notkestra{\ss}e 85, 22607 Hamburg, Germany}

\author{Maurits~W.~Haverkort}
\affiliation{Institute for Theoretical Physics, Heidelberg University, Philosophenweg 19, 69120 Heidelberg, Germany}

\author{Ilya~Elfimov}
\affiliation{Quantum Matter Institute, University of British Columbia, Vancouver, British Columbia, Canada V6T 1Z4}

\author{Gertrud~Zwicknagl}
\affiliation{Max Planck Institute for Chemical Physics of Solids, N{\"o}thnitzer Stra{\ss}e 40, 01187 Dresden, Germany}
\affiliation{Institute for Mathematical Physics, Technische Universit{\"a}t Braunschweig, D-38106 Braunschweig, Germany}

\author{Alexander~V.~Andreev}
\affiliation{Institute of Physics, Academy of Sciences of the Czech Republic, Na Slovance 1999/2, 182 21 Prague 8, Czech Republic}

\author{Ladislav~Havela}
\affiliation{Department of Condensed Matter Physics, Faculty of Mathematics and Physics, Charles University, Ke Karlovu 5, 121 16 Prague 2, Czech Republic}

\author{Mitchell~M.~Bordelon}
\affiliation{Los Alamos National Laboratory, Los Alamos, New Mexico 87545, USA}

\author{Eric~D.~Bauer}
\affiliation{Los Alamos National Laboratory, Los Alamos, New Mexico 87545, USA}

\author{Priscila~F.~S.~Rosa}
\affiliation{Los Alamos National Laboratory, Los Alamos, New Mexico 87545, USA}

\author{Andrea~Severing}
\affiliation{Max Planck Institute for Chemical Physics of Solids, N{\"o}thnitzer Stra{\ss}e 40, 01187 Dresden, Germany}
\affiliation{Institute of Physics II, University of Cologne, Z\"{u}lpicher Stra{\ss}e 77, 50937 Cologne, Germany}

\author{Liu~Hao~Tjeng}
\affiliation{Max Planck Institute for Chemical Physics of Solids, N{\"o}thnitzer Stra{\ss}e 40, 01187 Dresden, Germany}

\date{\today}

\begin{abstract}
We investigate the topological superconductor candidate UTe$_2$ using high-resolution valence-band resonant inelastic x-ray scattering at the U\,$M_{4,5}$-edges. We observe atomic-like low-energy excitations that support the correlated nature of this unconventional superconductor. These excitations originate from the U\,$5f^2$ configuration, which is unexpected since the short Te2-Te2 distances exclude Te2 being 2-. By utilizing the photoionization cross-section dependence of the photoemission spectra in combination with band structure calculations, we infer that the stabilization of the U\,$5f^2$ configuration is due to the U\,$6d$ bonding states in the U-dimers acting as a charge reservoir. Our results emphasize that the description of the physical properties should commence with a $5f^2$ \textit{ansatz}.
\end{abstract}



\maketitle


\section{Introduction}
UTe$_2$ is a recently-discovered odd-parity superconductor that emerged as a promising candidate for topological superconductivity. Since the discovery of superconductivity in 2019\,\cite{Ran2019,Aoki2019,Visser2019}, a plethora of experimental and theoretical studies has boosted our understanding of this intriguing material (see, e.g,\,\cite{Lewin2023} and references therein). Two central issues, however, remain unsettled even under ambient conditions. First, numerous proposals exist for the superconducting order parameter, which include either multicomponent\,\cite{Hayes2021,Ishizuka2021,Ishihara2023} or single-component representations\,\cite{Rosa2022,Suetsugu2023,Theuss2023,Lee2023}. Second, whether the local uranium multiplet configuration is U$^{3+}$ ($5f^{3}$) or U$^{4+}$ ($5f^{2}$) remains a matter of fierce debate\,\cite{Fujimori2019,Miao2020,Thomas2020,Fujimori2021,Shick2021,Aoki2022b,Liu2022,Wilhem2023}.

\begin{figure*}[t]
	\begin{center}
		\includegraphics[width=1.7\columnwidth,keepaspectratio]{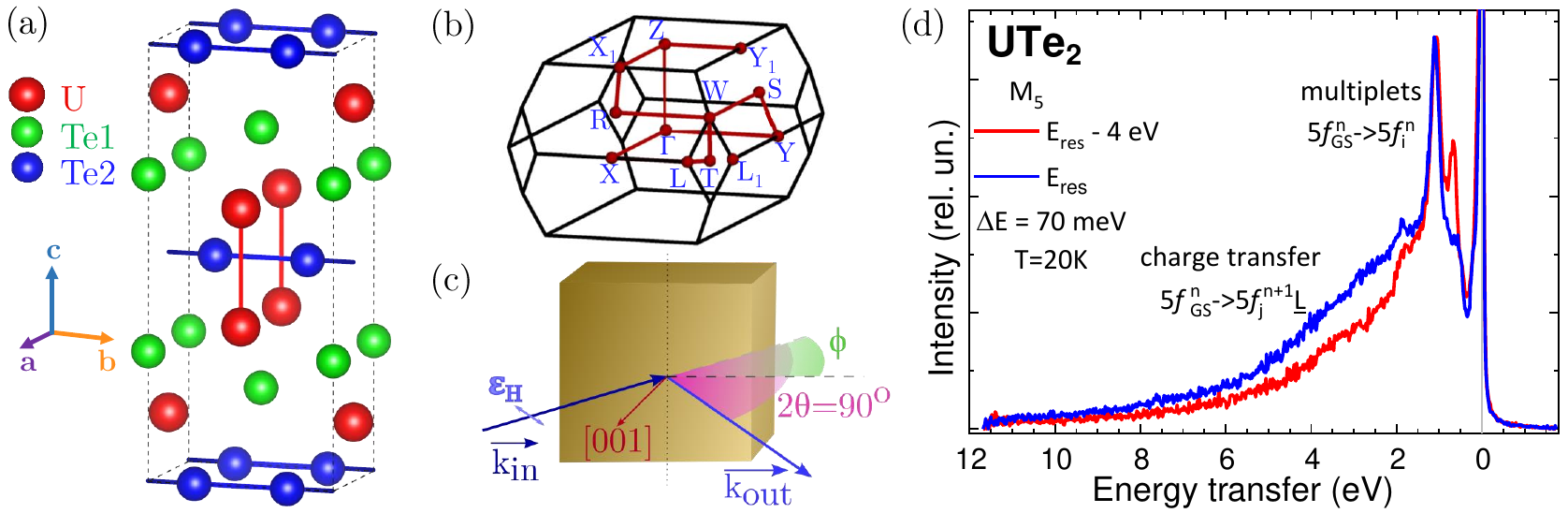}
		\vspace{-0.8cm}
	\end{center}
	\caption{Crystal structure of UTe$_2$ (a). First Brillouin zone of UTe$_2$ showing the main high-symmetry paths (b). Geometry of RIXS experiment, with scattering angle 2$\theta$, sample angle $\phi$, wave vector of incoming and outgoing x-rays $\vec{k}_{in,out}$, and $\epsilon_H$ horizontal polarization (c). $M_5$-edge valence band RIXS spectrum of UTe$_2$ in a large energy window showing, in addition to elastic scattering, multiplet and charge transfer excitations (d).}  
	\label{fig:structure}
\end{figure*}

\begin{figure}[b]
	\begin{center}
		\includegraphics[width=0.99\columnwidth,keepaspectratio]{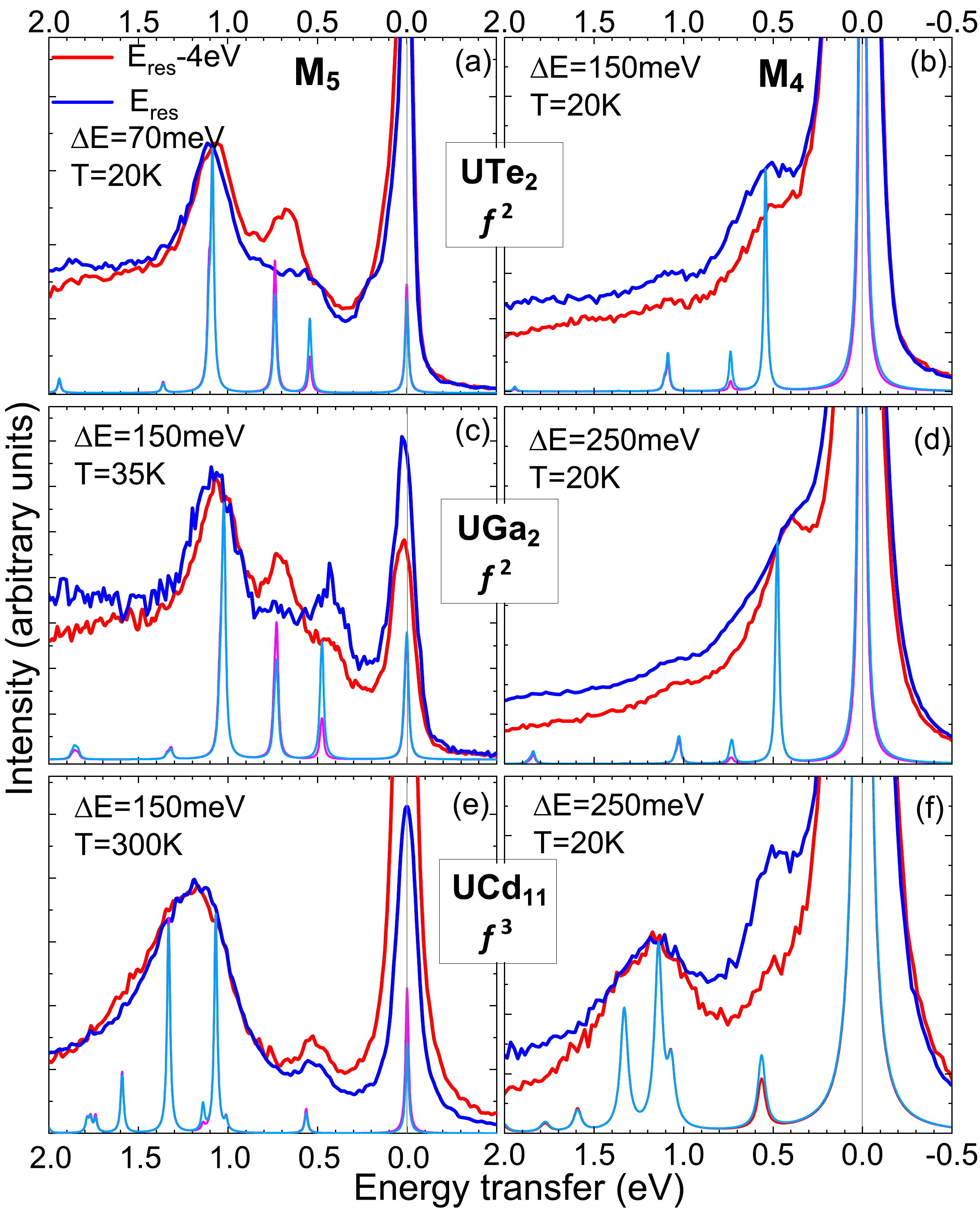}
		\vspace{-0.5cm}
	\end{center}
	\caption{(a) to (f): (a) to (f): Valence band $M$-edge RIXS spectra for two different incident photon energies, at resonance E$_{res}$ (blue, light blue) and at E$_{res}$\,-\,4eV (red, magenta), for UTe$_2$ (a) \& (b), UGa$_2$ (c) \& (d) and UCd$_{11}$ (e) \& (f) at the M$_5$- (left) and M$_4$-edge (right), respectively. The thicker lines represent the data, the thinner lines the atomic multiplet calculations (see text).}
	\label{fig:rixs}
\end{figure}

UTe$_2$ crystallizes in the orthorhombic $Immm$ structure (space group 71)\,\cite{Hutanu2020}, as depicted in Fig.\,\ref{fig:structure}\,(a) and (b), and exhibits superconductivity with a critical temperature ($T_c$) of 2.1\,K in the latest generation of samples grown through the molten salt technique\,\cite{Sakai2022}. The upper critical field is anisotropic and much higher than the Pauli limit expected for even-parity superconductors\cite{Lewin2023,Wu2023}. The NMR Knight-shift drop at $T_c$ is also much smaller than expected for an even-parity superconducting state\,\cite{Matsumura2023}. Importantly, the ground state of UTe$_2$ can be easily tuned through external parameters. Re-entrant superconductivity has been observed in magnetic fields applied either along $b$ or between the $b$ and $c$ axes\,\cite{Knebel2019,Ran2019b}. Hydrostatic pressure suppresses $T_c$ at first, but above a pressure of only 0.3~GPa, two superconducting transitions are identified. At even higher pressures antiferromagnetic order is observed\,\cite{Knafo2023}. More recent investigations show that UTe$_2$ undergoes a structural phase transition to a body centered tetragonal phase between 3 and 5\,GPa\,\cite{Ran2020,Thomas2020,Huston2022,Honda2023}.

A key challenge in understanding anisotropic properties, as well as their dependence on tuning parameters, is finding the appropriate \textit{ansatz} for describing such a complex $f$-electron system, situated at the border between localization and itineracy. Electronic structure investigations are essential to determine the extent to which local physics persists and to identify the electronic configuration that governs the unusual low-temperature properties. UTe$_2$ has been investigated by several groups using various experimental methods, but the interpretation of the data does not yield a uniform picture\,\cite{Fujimori2019,Miao2020,Thomas2020,Fujimori2021,Shick2021,Aoki2022b,Liu2022,Wilhem2023}.. 
 
Generally, uranium-based intermetallics are intermediate valent with a valence between 3+ and 4+ and, in the presence of electron correlations, it is important to know which configuration, the U\,$f^3$ or the U\,$f^2$, dominates and sets the stage for the ground state symmetry. Unfortunately,  discrepancies in determining the dominant $f$ configuration are common, as seen in, e.g., \cite{Jeffries2010,Fujimori2012,Wray2015,Booth2016,Sundermann2016,Kvashnina2017}. We believe this apparent contradiction arises from the challenge to quantitatively analyze the spectra. In general, this determination is difficult due to the complexity of the underlying many-body problem and uncertainties in the input model or input parameters. This issue is particularly serious when extracting information about the electronic states from fine variations in the energy positions or intensities of the spectral features. 

Here, we choose a spectroscopic method that provides a ``fingerprint", i.e., a distinctive set of strong spectral features if local $f$-electron physics is present and that are unique for any U\,5$f^n$ configuration\,\cite{Bright2023}. In our previous work\,\cite{Marino2023}, we demonstrated that a novel setup for high-resolution valence-band (VB) Resonant Inelastic X-ray Scattering (RIXS) at the U\,$M_5$-edge (3$d$\,$\rightarrow$ \,5$f$) in the tender x-ray regime offers sufficient resolution to reveal atomic-like low-energy excitations in UGa$_2$, a prototypical uranium intermetallic material. We successfully distinguished whether the main configuration is $5f^2$ or $5f^3$. The signal-to-background ratio of VB-RIXS at the U\,$M_5$-edge is high. In this study, we additionally utilize the U\,$M_4$-edge to provide yet another ``fingerprint" of localized 5$f$ configuration, further confirming the reliability of our method.

The U\,$M_{4,5}$-edge RIXS experiments were conducted at the Max-Planck IRIXS endstation of the P01 beamline at Petra III/DESY (Hamburg, Germany)\,\cite{Ketenoglu2015,Said2018,Gretarsson2020}. The resolution at the U\,M$_5$-edge (3.5\,keV) initially was 150\,meV and later improved to 70\,meV. At the U\,M$_4$-edge (3.7\,keV), the resolution progressed from 250\,meV to 150\,meV. The experimental set-up is depicted in Fig.\,\ref{fig:structure}\,(c). Spectra were gathered at various incident photon energies across the absorption edges to capitalize on the energy dependence of the cross-section. We will focus on the energies at the resonance maximum, E$_{res}$, and 4\,eV below the resonance, E$_{res}$\,-\,4\,eV.  Additional details about the experiment and sample synthesis can be found in the Appendix.

\section{Results}

Fig.\,\ref{fig:structure}\,(d) displays $M_5$-edge VB RIXS data of UTe$_2$ in an energy range up to 12\,eV for two incident energies, E$_{res}$ (blue curves) and E$_{res}$\,-\,4\,eV (red curves). In addition to the elastic peak, we observe sharp peaks that we attribute to U\,5$f$ atomic-like multiplet excitations from the ground state (GS) of the type 5$f^n_{GS}$\,$\rightarrow$\,5$f^n_i$, and a broad shoulder which we associate with charge transfer scattering reaching 5$f^{n+1}_j\underline{L}_j$ final states where \underline{L} denotes a ligand hole. The multiplet excitations determine the leading 5$f$ configuration in the mixed valent ground state as we will show below, and the charge transfer signal reflects the intermediate valent character of UTe$_2$. Fig.\,\ref{fig:rixs}\,(a)-(f) presents experimental M$_5$- and M$_4$-edge VB-RIXS spectra in a smaller energy window of -0.5 to 2\,eV of UGa$_2$, UCd$_{11}$, and UTe$_2$ for the same incident energies. UTe$_2$ data are compared to those from ferromagnetic UGa$_2$ ($T_C$\,=\,125\,K\,\cite{Lawson1985}), serving as a $5f^2$ reference material\,\cite{Marino2023}, and antiferromagnetic UCd$_{11}$ ($T_N$\,=\,5.3\,K\,\cite{Yamamoto2012}), serving as an $5f^3$ reference\,\cite{Booth2016,Amorese2020}. The $M_5$ spectra are normalized to the most intense inelastic peak just above 1\,eV energy transfer without any background subtraction. 

\begin{figure*}[t]
	\begin{center}
		\includegraphics[width=2.0\columnwidth]{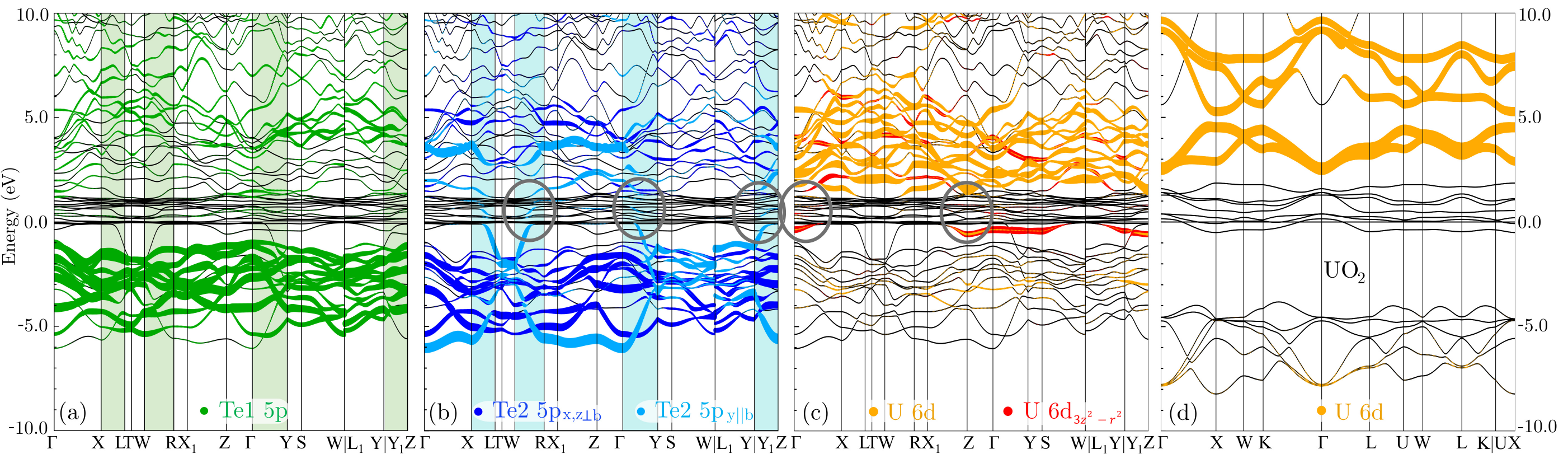}
		\vspace{-0.5cm}
	\end{center}
	\caption{Projected full relativistic band structure (DFT) of UTe$_2$ (a)-(c): 5$p$ bands of Te1 (green) in (a), Te2 (blue) with 5$p_{y\parallel b}$ states (light blue) in (b), and U 6$d$ bands (gold) and U 6$d_{3z^2-r^2}$ bands (red) in (c). Shaded regions represent the high-symmetry cuts along $b$. Band structure of UO$_2$ with U 6$d$ (gold) in (d). The gray ellipses in (b) and (c) highlight  regions of \textit{avoided crossing}.}
	\label{fig:TeBands}
\end{figure*}

\begin{figure}[]
	\begin{center}
		\includegraphics[width=0.85\columnwidth]{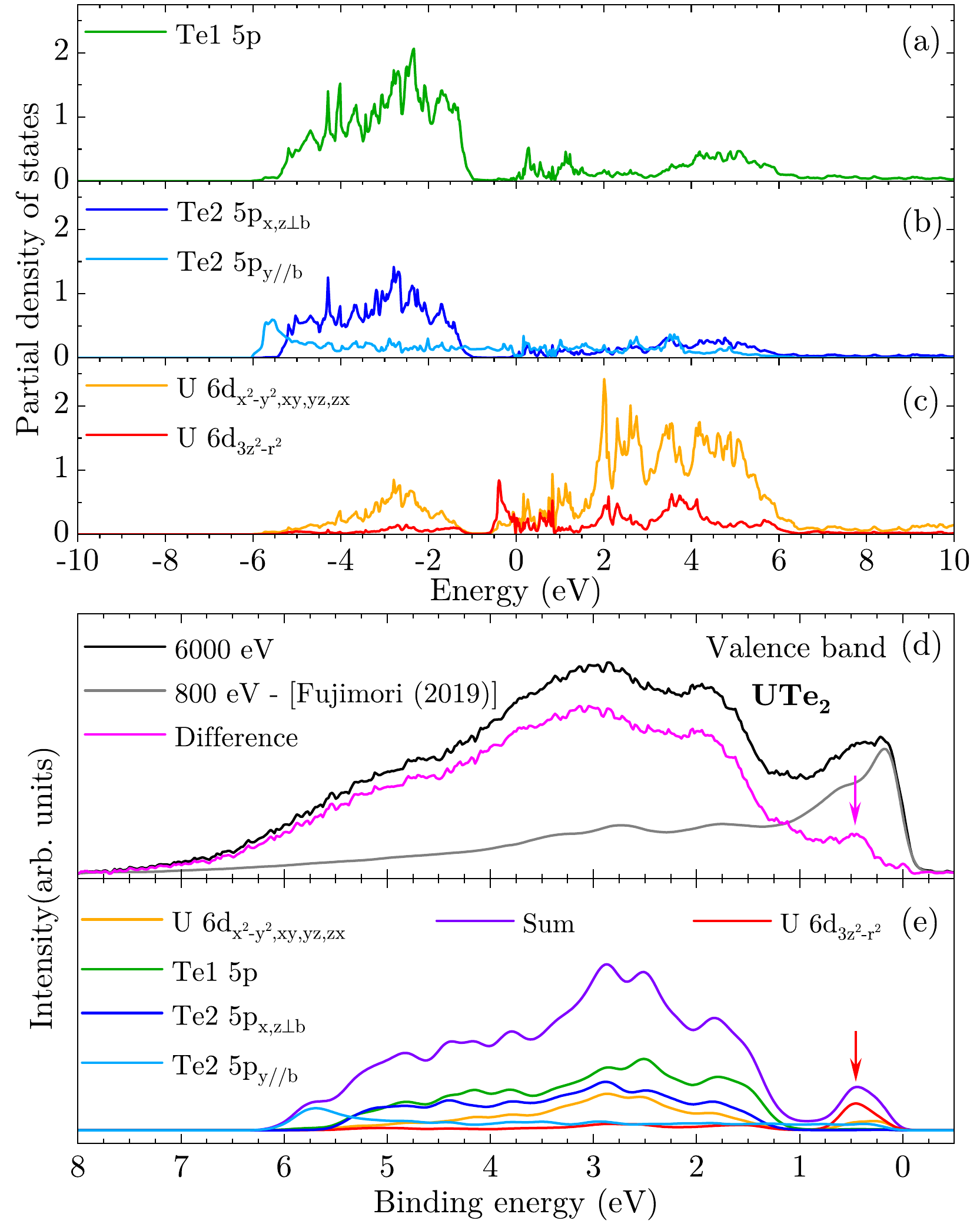}
		\vspace{-0.5cm}
	\end{center}
	\caption{(a)-(c) Partial density of states (DOS) corresponding to the band structure in Fig.\,\ref{fig:TeBands}. (d) Photoemission valence band data of UTe$_2$ measured with 6000\,eV (black) and data at 800\,eV (gray) as adapted from \cite{Fujimori2019}, and the difference plot (magenta) that no longer contains the 5$f$ states. The arrow highlights the shallow peak around 0.5\,eV. (e) Simulation of difference plot by adding up the respective partial DOSs from (a)-(c), weighted by their respective energy and shell-dependent photoionization cross-sections (see text). The arrow indicates the enhanced U\,6$d_{3z^2-r2}$ contribution at 0.5\,eV.}
	\label{fig:VB}
\end{figure}

Upon comparing the $M_5$ edges, we observe a prominent inelastic feature just above 1\,eV in all samples, which appears broader and at slightly higher energy transfer in UCd$_{11}$. The $M_5$-edge RIXS data of UGa$_2$ and UTe$_2$ reveal two additional excitations below the strongest feature, whereas the UCd$_{11}$ spectra in the lower energy region show only one less pronounced excitation. At the $M_4$-edge, the strong excitation above 1 eV persists only in the UCd$_{11}$ data, while it is strongly suppressed in UGa$_2$ and UTe$_2$. As demonstrated below, the intensity suppression around 1\,eV in the $M_4$ edge is a characteristic cross-section dependence for the U$^{4+}$\,5$f^2$ configuration.

RIXS spectra are simulated using full atomic multiplet calculations implemented in the $Quanty$ code \cite{Haverkort2016}, starting with the atomic input values from the Atomic Structure Code by R. D. Cowan\,\cite{Cowan1981}. These values are reduced to account for covalence effects and configuration interaction, which are not included in the Hartree-Fock scheme\,\cite{Tanaka1994,Groot2008,Agrestini2017}. Further information about the simulation can be found in the Appendix.

Multiplet excitations are configuration specific so that they show at the same energy in the $M_5$ and $M_4$-edge RIXS spectra. However, cross-sections may vary from edge to edge due to different quantum numbers involved. Consequently, the spectral weights of the respective multiplet excitations can differ significantly at the $M_5$- or $M_4$-edge. Indeed, simulations based on a U\,5$f^2$ configuration capture the suppression of spectral weight around 1\,eV in the $M_4$-edge data compared to $M_5$ for UGa$_2$ and UTe$_2$ (see simulations in Fig.\,\ref{fig:rixs}\,(a)\&(b) and (c)\&(d)). Conversely, a simulation based on the 5$f^3$ configuration would yield much more intensity in this energy range at the $M_4$-edge, as evident in Fig.\,\ref{fig:rixs}\,(e)\&(f) for UCd$_{11}$.
 
The good agreement between experimental and simulated cross-section dependence not only confirms the U\,5$f^2$ analysis of UGa$_2$ in Ref.\,\cite{Marino2023}, it also supports the assumption of Ref.\,\cite{Booth2016,Amorese2020} that  UCd$_{11}$ is a 5$f^3$ compound. In contrast, UTe$_2$, akin to UGa$_2$, exhibits suppressed spectral weight of the multiplet structure around 1\,eV when comparing the $M_5$ edge to the $M_4$-edge. Hence, we conclude that the $M$-edge RIXS spectra of UTe$_2$ resemble the multiplet structure of the $5f^2$ configuration, supporting the (less intense) $O_{4,5}$-edge RIXS data and analysis by Liu \textit{et al}.\,\cite{Liu2022}.

\section{Discussion}

The presence of dominant local multiplet excitations in the M$_{4,5}$-edge RIXS spectra of UTe$_2$ is a direct signature of  electronic correlations in the U\,$5f$ shell. However, the 5$f^2$ character of the main local uranium configuration raises questions from a chemistry point of view because it implies that the U is formally $4+$, suggesting all Te ions in UTe$_2$ should be formally $2-$. A Te$^{2-}$ ion is a notably large anion with an ionic radius of 2.21\,\AA\,\cite{Jovic1998}. For comparison, the Te-Te distance in EuTe (Eu$^{2+}$, Te$^{2-}$) is as large as 4.67\,\AA\,\cite{Lafrentz2010}. Examination of the Te-Te distances in UTe$_2$ reveals that only the Te1-Te1 distances of 4.123\,\AA\, are compatible with such a large ion. Te2 ions form chains along the $b$ direction with Te2-Te2 distances of only 3.0355\,\AA\,\cite{Hutanu2020} (see Fig.\,\ref{fig:structure}), which is too short to support a divalent state. These considerations were already put forward by St{\"o}we \cite{Stowe1997}, who proposed the \textit{formal} chemical charge state U$^{3+}$Te$_1^{2-}$Te$_2^{1-}$ if integer valences were to be used.

We performed Density Functional Theory (DFT) calculations for UTe$_2$ to investigate the U and Te valences. The calculations were performed using the full potential local-orbital code (FPLO)\,\cite{Koepernik1990}, employing a local density approximation (LDA) (functional from Perdew and Wang\,\cite{Perdew1992}) in a full-relativistic approach that includes the spin orbit splitting (see Appendix for further details). In Fig.\,\ref{fig:TeBands} and Fig.\,\ref{fig:VB}, we highlight the 5$p$ bands of Te1 (green) and Te2 (dark/light blue), and the 6$d$ of U (gold/red). While the results for the correlated U\,5$f$ states require caution, the energy positions and dispersions of the broader Te $5p$ and U 6$d$ bands can be reasonably trusted. In Fig.\,\ref{fig:TeBands}\,(a), all thick Te1\,5$p$ bands lie 2\,eV below the Fermi level, consistent with the formal charge of $2-$. In contrast, Fig.\,\ref{fig:TeBands}\,(b) shows that parts of the Te2\,5$p$ bands cross the Fermi level, exhibiting strong dispersion in the shaded regions that are cuts along the $b$ direction (see 1st Brilloin zone in Fig.\,\ref{fig:structure}\,(b)), in agreement with ARPES measurements along the $\overline{\Gamma} - \overline{Y}$ cuts\,\cite{Miao2020}. These strongly dispersing bands are the Te2\,5$p_{y\parallel b}$ bands (highlighted in light blue), forming bonding and anti-bonding bands along the Te2 chains. Roughly half of these bands are below the Fermi level, and the other half is above $E_F$, as shown by the Te2\,5$p_{y\parallel b}$ partial density of states (pDOS) in light blue in Fig.\,\ref{fig:VB}\,(b). These results support the notion of Te2 being $1-$ and not $2-$, meaning that one electron is donated from the Te2\,5$p_{y\parallel b}$ states to U. 

Experimentally, though, we find the formal U\,5$f^2$ configuration in UTe$_2$. Hence, the donated electron must go somewhere else. Therefore, we explore the role of the U\,6$d$ states.  For this, we compare the band structure of UTe$_2$ to that of UO$_2$, a well-studied semiconductor with a recognized U$^{4+}$\,5$f^2$ configuration (see Fig. \ref{fig:TeBands}\,(c)\,\& \,(d)). In UO$_2$, the $6d$ bands (gold) are situated well above the Fermi level, with very little $6d$ character below, confirming the local configuration of U in UO$_2$ as 5$f^2$6$d^0$. In contrast, UTe$_2$ exhibits the empty 6$d$ bands (gold/red) much closer to the Fermi level, with substantial amount of $6d$ character below the Fermi level. Notably, a filled 6$d$ band (red) at around -0.5\,eV has $3z^2$\,-\,$r^2$ character and is the bonding part of the two 6$d$ orbitals pointing towards each other along the U dimer parallel to the $c$ axis. We conclude, the U\,6$d_{3z^2-r^2}$ absorbs the extra electron from Te2. Interestingly, this filled 6$d_{3z^2-r^2}$ band is non-bonding concerning the Te1 and Te2\,5$p$ bands, as evidenced from the absence of common characters or features in the bands and pDOS at -0.5\,eV in Fig.\,\ref{fig:TeBands}\,(a)-(c)) Fig.\,\ref{fig:VB}\,(a)-(c). 

The partial filling of the 6$d_{3z^2-r^2}$ at -0.5\,eV as derived from band structure is supported experimentally by photoelectron spectrocopy measurements of the valence band (VB) with hard (HAXPES) and soft x-rays (PES). In HAXPES, the cross-section of the U\,5$f$ states is strongly reduced with respect to the U\,6$d$ and Te\,5$p$ states, whereas the U\,5$f$ states dominate the intensity in the PES spectra \,\cite{Trzhaskovskaya2001,Trzhaskovskaya2002,Trzhaskovskaya2006, Rosner2009,Takegami2019,Takegami2022,Altendorf2023}. 
Hence we can extract the non-$f$ states using this strong photon energy dependence of the cross-sections. Fig.\,\ref{fig:VB}\,(d) displays the HAXPES VB data taken with 6000\,eV x-rays (black curve) together with the cross-section corrected PES spectrum taken with 800\,eV photons by Fujimori \textit{et al.}\,\cite{Fujimori2019} (gray curve). A detailed description of the HAXPES experimental set-up\,\cite{Schlueter2019}, adaption of the Fujimori data, and cross-section correction is given in the Appendix. The difference spectrum, see magenta curve in Fig.\,\ref{fig:VB}\,(d), no longer contains contribution from the U\,5$f$ states. This spectrum is compared to the sum of the cross-section weighted pDOS of the U\,6$d$ (gold/red), Te1 (green) and Te2 (dark/light blue) 5$p$, as shown in Fig.\,\ref{fig:VB}\,(e). The agreement between experiment and calculations is good, notably featuring a peak at 0.5 eV binding energy in the experiment (indicated by the magenta arrow) that matches very well the peak in the 6$d_{3z^2-r2}$ pDOS (red arrow).

We note that the Te2\,5$p_{y\parallel b}$ bands are essentially non-bonding to the other Te $5p$ and U $6d$ bands. This is evident from the lack of common characters or features in Fig.\,\ref{fig:TeBands}\,(a)-(c) and Fig.\,\ref{fig:VB}\,(a)-(c). Thus, the Te2\,5$p_{y\parallel b}$ bands form a subsystem well-separated from the rest. The charge transfer of about one electron from Te2\,5$p_{y\parallel b}$ to U is thus independent of the charge transfer that also occurs between the Te1\,5$p$, Te2\,5$p_{x,z\perp b }$ to the U 6$d$ bands due to hybridization.

Finally, we explore the role of the 5$f$ states. In the calculations, the U\,5$f$ states are positioned around the Fermi level. Upon examining Fig.\,\ref{fig:TeBands}\,(a)-(c), a noticeable avoidance of crossing is observed between these U\,5$f$ states (black lines) and the Te2\,5$p_{y\parallel b}$ states, and (at different regions in k-space) even more so with the 6$d_{3z^2-r^2}$ states (see gray ellipses in Fig.\,\ref{fig:TeBands}\,(b)and (c)), showing hybridization. Therefore, the low-energy Hamiltonian of UTe$_2$ should commence with the local 5$f^2$ configuration found in the RIXS experiment, the 5$p_{y\parallel b}$ states of the Te2 chains and the bonding states of the 6$d_{3z^2-r2}$ orbitals in the U-dimer.

\section{CONCLUSION}
In summary, high-resolution M${4,5}$-edge RIXS data of UTe$_2$ reveal atomic-like U\,$5f$-$5f$ excitations from the U\,5$f^2$ configuration, confirming the correlated nature of UTe$_2$ despite strong covalence and settling current debates concerning the dominating valence. The puzzle of the short Te2-Te2 distances, incompatible with formal U$^{4+}$ valence, is resolved through band structure calculations and photon energy-dependent photoemission, showing charge transfer from 5$p_{y\parallel b}$ states of the Te2 chain is directed not to the U\,5$f$ but to the bonding state of the U\,6$d_{3z^2-r2}$ orbitals of the U-dimer. Both hybridize with the U\,5$f$ states, so that the description of the physical properties of UTe$_2$ should include these two bands together with a $5f^2$ \textit{ansatz}. We thus propose a scenario of partial entanglement of 5$f$ states, involving a 5$f^2$ configuration and some extra 5$f$ electrons, likely forming bands or potentially participating in a Kondoesque manner.
	
\section{Acknowledgment}
All authors thank Ulrich Burkhardt, Katharina Höfer, Anna Melendez-Sans and Simone Altendorf for assistance with the sample preparation. A.S acknowledges support from the German Research Foundation (DFG) - grant N$^{\circ}$ 387555779 and M.H. from the Bundesministerium für Bildung und Forschung (BMBF) - grant 13K22XXB. All authors acknowledge DESY (Hamburg, Germany), a member of the Helmholtz Association HGF, for the provision of experimental facilities. Work at Los Alamos National Laboratory was performed under the auspices of the U.S. Department of Energy, Office of Basic Energy Sciences, Division of Materials Science and Engineering under project “Quantum Fluctuations in Narrow-Band Systems”. M.M.B. acknowledges support from the Los Alamos Laboratory Directed Research and Development program. A.V.A. and L.H. benefited from financial support of the Czech Science Foundation - project N$^{\circ}$ 21-09766S.

\section{Appendix}
\subsection{Samples} Plate-like single crystalline samples of UTe$_2$ were synthesized using two methods: the chemical vapor transfer (CVT) method as described in \cite{Rosa2022} and the molten salt flux (MSF) technique as described in \cite{Sakai2022}. CVT samples show a superconducting transition temperature around 1.9~K, whereas MSF samples become superconducting at 2.1~K. Laue backscattering patterns were performed to confirm that the $c$ axis is perpendicular to the plate. UCd$_{11}$ single crystals (P$m\overline{3}m$ (221) cubic space group) were grown with the metallic flux growth technique\,\cite{Canfield1992} and UGa$_2$ single crystals (P6/$mmm$ (191) hexagonal space group) were grown with the Czochralski method\,\cite{Kolomiets2015}. Their crystalline phase and stoichiometry were determined through single crystal/powder X-ray diffraction and energy-dispersive X-ray spectroscopy. 

\subsection{RIXS experimental set-up} The design of the Max-Planck IRIXS endstation of the P01 beamline at Petra III/DESY (Hamburg, Germany)\,\cite{Gretarsson2020} is based on a hard x-ray setup. IRIXS covers the tender energy range of 2.4 to 4.0\,keV, thus making cleaving \textit{in-situ} obsolete. High resolution at the U $M$-edges was achieved with a diced quartz wafer using at first the 112 reflection and, for further improvement, the 003. The quartz waver was pressed and glued onto a concave Si lens to form an analyzer crystal\,\cite{Ketenoglu2015,Said2018}. In the present experiments, the resolution at the U\,M$_5$-edge (3.5\,keV) initially was 150\,meV and then, with the 003 reflection, improved to 70\,meV, while at the U\,M$_4$-edge (3.7\,keV) the resolution improved from 250\,meV to 150\,meV. The incoming beam is horizontally polarized and the scattering angle 2$\theta$ was set to 90$^{\circ}$ with a sample angle $\phi$\,=\,45$^{\circ}$ so that the [001] direction of all samples was parallel to the momentum transfer. Spectra were collected at several incident energies to exploit the energy dependence of the cross-section. The experiments were conduced in vacuum with a cryostat with a base pressure around high 10$^{-8}$\,mbar. For energy alignment, a carbon tape was measured regularly between the RIXS scans of the samples.

For the RIXS experiments UTe$_2$ crystals were carefully selected using macroscopic techniques. All samples were aligned and mounted with the $[001]$ direction normal to the holder surface, using the Laue diffraction method. The top surfaces of the mounted UTe$_2$ crystals were polished inside a glove-box (initial O$_2$ and H$_2$O concentration $<$ 1 ppm). Subsequently, they were transferred in a vacuum suitcase into an ultra-high vacuum system (low 10$^{-10}$\,mbar regime) with a molecular beam epitaxy chamber where a 45\,nm-thick capping layer of Cr was deposited.

\subsection{Simulation} The full multiplet simulations with the Quanty code\,\cite{Haverkort2016} take into account the intermediate state, spin-orbit coupling (SOC), and Coulomb repulsion between the 3$d$-5$f$ and 5$f$-5$f$ electrons. The crystal-field splitting of the Hund's rule ground state is considered only for UGa$_2$. This consideration arises from UGa$_2$ being the only compound in this study for which the RIXS data exhibit an orientation dependence\,\cite{Marino2023} i.e. where the anisotropic charge density of the ground state impacts the inelastic intensity ratios. The splitting in the multiplet excitations due to the crystal field, however, cannot be resolved with our present resolution. The atomic input values from the Atomic Structure Code by R. D. Cowan\,\cite{Cowan1981} are reduced to account for covalence effects and configuration interaction, which are not included in the Hartree-Fock scheme\,\cite{Tanaka1994,Groot2008,Agrestini2017}. The non-monopole 5$f$-5$f$ and  non-monopole 3$d$-5$f$ Slater integrals, as well as the spin orbit coupling (SOC) from the Cowan code are reduced by multiplying with the factors $R_{ff}$, $R_{fc}$, and $R_{SOC}$, respectively (see Table\,I).  These values were tuned to best describe the energy positions of the multiplets.

\begin{table}[h]
	\centering
	\caption{Reduction factors (in \%) by which the Slater integrals from the Cowan code \cite{Cowan1981} were multiplied with in the simulations.}
	\begin{tabular}{ccccc}
		Compound   &5$f^n$&$R_{ff}$ & $R_{fc}$ & $R_{SOC}$ \\ \hline
		UGa$_2$    &5$f^2$& 55(2)   & 60(20)    & 87(2) \\
		UCd$_{11}$ &5$f^3$& 52(2)   & 80(20)    & 89(2) \\
		UTe$_2$    &5$f^2$& 65(2)   & 80(20)    & 89(2) \\
	\end{tabular}
\end{table}

\subsection{Band structure} The electronic band structure calculations were performed using full potential local-orbital code (FPLO)\,\cite{Koepernik1990}, employing a local density approximation (LDA) (functional from Perdew and Wang\,\cite{Perdew1992}) in a full-relativistic approach that includes the spin orbit splitting. The solutions of the Kohn-Sham Dirac equations  were projected back into the atomic-non coupled $|nlm \rangle$ bases for clarity of interpretation. The Brillouin zones were sampled by a well-converged regular 15x15x15 $k$-point mesh in the full zone, and approximately one energy point every 20\,meV was used to calculate the band plots. The lattice parameters used for UTe$_2$ were $a$\,=\,4.18\,\AA, $b$\,=\,6.16\,\AA, $c$\,=\,14.02\,\AA\,\cite{Rosa2022} and $a$\,=\,5.469\,\AA \,for UO$_2$\,\cite{Momin1991}.

Figure\,S1 shows the calculation in a scalar relativist approach i.e. without spin orbit coupling. In this simpler calculation the dispersion of the Te2\,5$p_{y\parallel b}$, the partially filled 6$d_{3z^2-r^2}$ band, and the avoided crossings with the 5$f$ bands become even more obvious, showing that our calculations and conclusions are robust.

\begin{figure*}[t]
	\begin{center}
		\includegraphics[width=1.99\columnwidth]{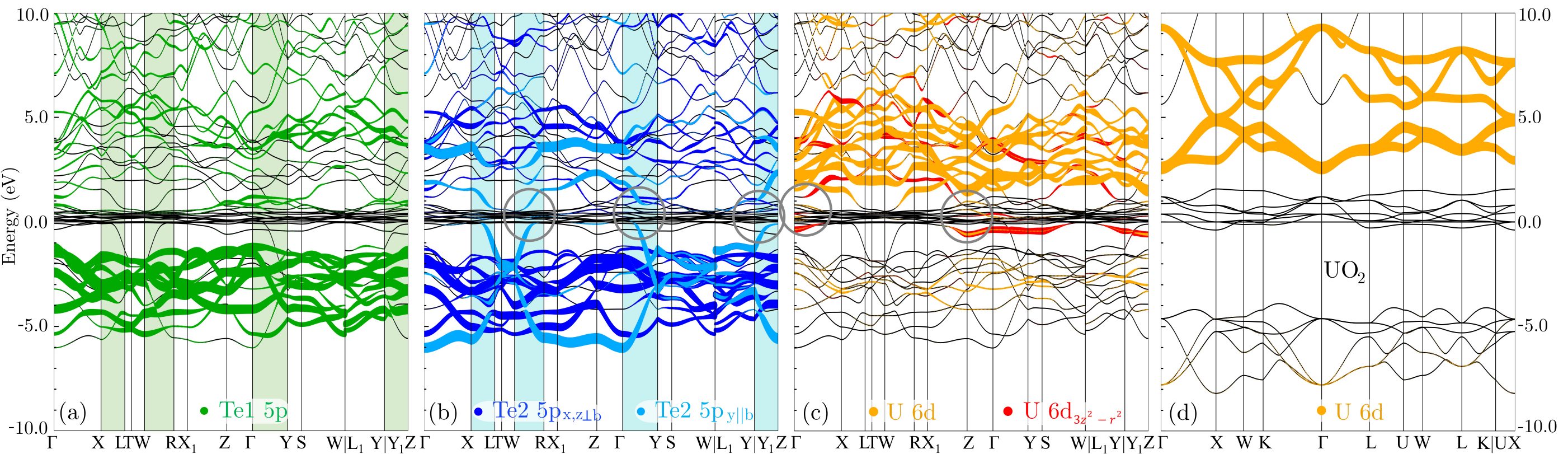}
		\vspace{-0.2cm}
	\end{center}
	\caption{FIG.\,S1\,\, Projected scalar relativistic band structure (DFT) of UTe$_2$ (a)-(c): 5$p$ bands of Te1 (green) in (a), Te2 (blue) with 5$p_{y\parallel b}$ states (light blue) in (b), and U 6$d$ bands (gold) and U 6$d_{3z^2-r^2}$ bands (red) in (c). Shaded regions represent the high-symmetry cuts along $b$. Band structure of UO$_2$ with U 6$d$ (gold) in (d). The gray circles in panels (b) and (c) highlight regions of \textit{avoided crossing}.}
\end{figure*}

\subsection{HAXPES experimental set-up} 
The  HAXPES  experiments  of the UTe$_2$ valence band were conducted  at  the P22 beamline at Petra III/DESY (Hamburg, Germany)\,\cite{Schlueter2019}. To determine the Fermi level, the valence band (VB) signal of a Au film was measured, setting our incident  photon  energy at  6001.5\,eV, with an  overall  instrumental  resolution  of  233\,meV. The emitted  photoelectrons were collected using a SPECS Phoibos 225HV analyzer in the horizontal plane at 90$^{\circ}$ from the incoming beam, with the sample emission angle set at 45$^{\circ}$. Clean surfaces were achieved by knife-cleaving the samples in an adjacent chamber at base pressure of 10$^{-10}$\,mbar before transferring them into the main chamber. The experiments were performed at $T$\,=\,50\,K. The VB measured with HAXPES is presented in Fig.\,4\,(d) as the black curve and is compared with data obtained by Fujimori \textit{et al}.\,\cite{Fujimori2019} using 800\,eV photons (gray curve). To facilitate comparison, we broadened the 800\,eV spectrum, taken with 140 meV resolution, by an additional 180\,meV to achieve the same overall resolution as for the 6000\,eV data. Exploiting the strong photon energy dependence of the photoionization cross-sections of the U\,6$d$ and Te\,$5p$ states relative to that of the U\,5$f$ states, we distinguish their contributions to the VB: 
$\sigma^{800}_{U6d}$/$\sigma^{800}_{U5f}$\,=\,0.10 and $\sigma^{800}_{Te5p}$/$\sigma^{800}_{U5f}$\,=\,0.14 at h$\nu$\,=\,800\,eV, and $\sigma^{6000}_{U6d}$/$\sigma^{6000}_{U5f}$\,=\,0.73 and $\sigma^{6000}_{Te5p}$/$\sigma^{6000}_{U5f}$\,=\,0.61 at h$\nu$\,=\,6000\,eV.


\subsubsection{Cross-section analysis}
In the following, we demonstrate the process of performing the cross-section anaylsis\,\cite{Trzhaskovskaya2001,Trzhaskovskaya2002,Trzhaskovskaya2006, Rosner2009,Takegami2019,Takegami2022,Altendorf2023}. This anaylsis is aimed at eliminating the U\,5$f$ states in the valence band HAXPES data by utilizing the soft X-ray data published by Fujimori \textit{et al.}\,\cite{Fujimori2019}. 

\begin{itemize}
    \item HPES(E) = HAXPES spectrum of UTe$_2$ taken at PETRA-III with 6000 eV photons;
    \item SPES(E)  = SXPES spectrum of UTe$_2$ taken by S. Fujimori with 800 eV photons after broadening and energy alignment;
    \item $A$ = unknown experimental factor for HAXPES experiment at PETRA-III;
    \item $B$ = unknown experimental factor for SXPES experiment by S. Fujimori;\\ \\
    (photon flux, analyzer/detector efficiency, measurement time)
    \item pDOS$_{U5f}$(E)  = U\,5$f$ partial density of states;
    \item pDOS$_{U6d}$(E)  = U\,6$d$ partial density of states;
    \item pDOS$_{Te5p}$(E) = Te\,5$p$ partial density of states;
    \item $\sigma^{6000}_{U5f}$   = U\,5$f$ photo-ionization cross-section (per electron) at h$\nu$\,=\,6000\,eV;
    \item $\sigma^{6000}_{U6d}$   = U\,6$d$ photo-ionization cross-section (per electron) at h$\nu$\,=\,6000\,eV;
    \item $\sigma^{6000}_{Te5p}$ = Te\,5$p$ photo-ionization cross-section (per electron) at h$\nu$\,=\,6000\,eV;
    \item $\sigma^{800}_{U5f}$   = U\,5$f$ photo-ionization cross-section (per electron) at h$\nu$\,=\,800\,eV;
    \item $\sigma^{800}_{U6d}$  = U\,6$d$ photo-ionization cross-section (per electron) at h$\nu$\,=\,800\,eV;
    \item $\sigma^{800}_{Te5p}$ = Te\,5$p$ photo-ionization cross-section (per electron) at h$\nu$\,=\,800\,eV.
\end{itemize}

The intensity of the valence band HAXPES and SXPES data can be written in terms of the sums of the contributing partial density of states times the respective cross-sections:

\begin{equation}
\begin{split}
HPES(E)&=A\cdot(\sigma^{6000}_{U5f} \cdot pDOS_{U5f}(E)+\sigma^{6000}_{U6d}\cdot pDOS_{U6d(E)}\\&+\sigma^{6000}_{Te5p}\cdot pDOS_{Te5p}(E))
\end{split}
\end{equation}

\begin{equation}
\begin{split}
SPES(E)&=B\cdot(\sigma^{800}_{U5f} \cdot pDOS_{U5f}(E)+\sigma^{800}_{U6d}\cdot pDOS_{U6d(E)}\\&+\sigma^{800}_{Te5p}\cdot pDOS_{Te5p}(E))
\end{split}
\end{equation}



Division by $\sigma^{6000}_{U5f}$\,$\cdot$\,$A$ and $\sigma^{800}_{U5f}$\,$\cdot$\,$B$, respectively, and we obtain:

\begin{equation}
\begin{split}
\frac{HPES(E)}{\sigma^{6000}_{U5f}\cdot A}&=pDOS_{U5f}(E)+\frac{\sigma^{6000}_{U6d}}{\sigma^{6000}_{U5f}}\cdot pDOS_{U6d}(E)\\&+\frac{\sigma^{6000}_{Te5p}}{\sigma^{6000}_{U5f}}\cdot pDOS_{Te5p}(E) 
\end{split}
\end{equation}

\begin{equation}
\begin{split}
\frac{SPES(E)}{\sigma^{800}_{U5f}\cdot B}&=pDOS_{U5f}(E)+\frac{\sigma^{800}_{U6d}}{\sigma^{800}_{U5f}}\cdot pDOS_{U6d}(E)\\&+\frac{\sigma^{800}_{Te5p}}{\sigma^{800}_{U5f}}\cdot pDOS_{Te5p}(E) 
\end{split}
\end{equation}

Subtracting the two equations, we obtain:

\begin{equation}
	\begin{split}
		\frac{HPES(E)}{\sigma^{6000}_{U5f}\cdot A}-\frac{SPES(E)}{\sigma^{800}_{U5f}\cdot B}=\\
		(\frac{\sigma^{6000}_{U6d}}{\sigma^{6000}_{U5f}}-\frac{\sigma^{800}_{U6d}}{\sigma^{800}_{U5f}}) \cdot pDOS_{U6d}(E)+\\
		(\frac{\sigma^{6000}_{Te5p}}{\sigma^{6000}_{U5f}}-\frac{\sigma^{800}_{Te5p}}{\sigma^{800}_{U5f}}) \cdot pDOS_{Te5p}(E)  
	\end{split}
\end{equation}

This difference no longer contains the 5$f$ states. The term on the left of the equal sign is the difference plot in Fig.\,4d (magenta line), with $A$ and $B$ as adjustable parameters, while the term on the right of the equal sign is the sum plotted in Fig\,4e (purple line) of the manuscript.


%
\end{document}